\pdfoutput=1

\RequirePackage{fix-cm}

\documentclass[onecolumn,numbers,sort&compress]{svjour3}

\smartqed

\usepackage{graphicx}
\usepackage{amssymb,amsmath}
\usepackage{txfonts}
\usepackage{natbib}
\usepackage{hyperref}

\hypersetup{pdftitle = The title of my PDF, pdfauthor = My name, pdfsubject= The subject, pdfkeywords = keyword1 keyword2 keyword3}
\hypersetup{colorlinks = true, linkcolor = blue, anchorcolor = red, citecolor = blue, filecolor = red, pagecolor = red, urlcolor = blue}

\journalname{Few-Body Systems}

\newcommand{\mathz}{\ooalign{$z$\cr\hfil\rule[.5ex]{.2em}{.06ex}\hfil\cr}}

\begin{document}

\title{Investigating the basins of convergence in the circular Sitnikov three-body problem with non-spherical primaries}

\author{Euaggelos E. Zotos \and Md Sanam Suraj \and Rajiv Aggarwal \and Satyendra Kumar Satya}

\institute{Euaggelos E. Zotos: \at
             Department of Physics, School of Science, \\
             Aristotle University of Thessaloniki, \\
             GR-541 24, Thessaloniki, \\
             Greece \\
             \email{\url{evzotos@physics.auth.gr}}
         \and
           Md Sanam Suraj: \at
             Department of Mathematics, Sri Aurobindo College, \\
              University of Delhi, Delhi, \\
              India \\
             \email{\url{mdsanamsuraj@gmail.com}}
         \and
           Rajiv Aggarwal: \at
             Department of Mathematics, Sri Aurobindo College, \\
             University of Delhi, Delhi, \\
             India \\
             \email{\url{rajiv_agg1973@yahoo.com}}
         \and
           Satyendra Kumar Satya: \at
             Department of Mathematics, LNJ College, \\
             Madhubani, Bihar, \\
             India
             \email{\url{sksatya09@gmail.com}}
}

\date{Received: 9 March 2018 / Accepted: 27 April 2018}

\titlerunning{Basins of convergence in the Sitnikov three-body problem}

\authorrunning{E.E. Zotos et al.}

\maketitle

\begin{abstract}

In this work we numerically explore the Newton-Raphson basins of convergence, related to the equilibrium points, in the Sitnikov three-body problem with non-spherical primaries. The evolution of the position of the roots is determined, as a function of the value of the oblateness coefficient. The attracting regions, on several types of two dimensional planes, are revealed by using the classical Newton-Raphson iterative method. We perform a systematic and thorough investigation in an attempt to understand how the oblateness coefficient affects the geometry as well as the overall properties of the convergence regions. The basins of convergence are also related with the required number of iterations and also with the corresponding probability distributions.


\end{abstract}

\section{Introduction}
\label{intro}

The restricted problem of three bodies has fascinated many scientists and researchers from Newton to the present and, it is the most celebrated of all dynamical problems. The Sitnikov problem is a special case of the restricted three-body problem where the test particle, of mass $m$, oscillates along the $z-$axis perpendicular to the configuration $(x,y)$ plane, in which two equally massed primary bodies, with masses $m_1$ and $m_2$, move in circular or elliptic orbits with common barycenter i.e. the axes origin $O$. Actually, in the simple case, where the primaries move in circular orbits $(e=0)$ the problem is also known as the MacMillan problem \cite{McM11}. It was \cite{P07}, who originally introduced this dynamical model when the primaries are moving in circular orbit. Furthermore, \cite{McM11} shown that the exact solution can be evince by Jacobi elliptic integral which have been also discussed in detail by \cite{S65}. The first qualitative results for special orbit have been studied by \cite{S60}, while \cite{M73} discussed the problem in the same vein.

In the past few decades, the Sitnikov problem has been studied by many scientists including various perturbations (e.g., \cite{C99,CL00,D93,DS97,F02,JE01,JP97,LS90,PM87,W93}). An analytical approach to the elliptic Sitnikov three-body problem is introduced by \cite{H92,H09}. Moreover, \cite{F03} started similarly from the equation of motion but he had applied a low order expansion to the problem. The presented solution is valid for small bounded oscillations in cases of moderate primary eccentricities. In addition, \cite{HL05} have presented the high order perturbation analysis of the Sitnikov problem using Floquet theory to derive the solutions of the linearized equation up to 17-th order in eccentricity.

A large number of scientists devoted their effort to study the Sitnikov three-body problem including various perturbation, such as: the effects of radiation (e.g., \cite{PK06}), the prolateness of the primaries (e.g., \cite{DKMP12}), the oblateness of the primaries (e.g., \cite{RGH15}). The families of periodic orbits and the corresponding bifurcations in the Sitnikov three-body problem are also discussed by many authors (e.g., \cite{BLO94,KPP08,Per07}). Furthermore, \cite{SBV07} have studied the stability of motion in the Sitnikov three-body problem. In particular, they located evidently infinite sequence of stability intervals on the $z-$axis and they also observed that as we move far from the primaries the width of these intervals tends asymptotically, while on the other hand the distance between them decreases. Additionally, in the Sitnikov problem, where the third mass is not negligible, they observed that as the value of third mass increases, the regions of bounded motion steadily grow and the third mass oscillates with larger and larger amplitudes along the $z-$direction.

The study of the basins of convergence, associated with the equilibrium points, is really very important since it reveals the most intrinsic properties of the dynamical system. Recently, many authors have studied the Newton-Raphson basins of convergence associated with libration points in the restricted three, four or even five-body problems, including various types of perturbations (e.g., \cite{SAA17,SAP17,Z16,Z17,ZS18}). On this basis, it is very interesting to introduce these ideas in the circular Sitnikov three-body problem with spheroidal primaries. In the present study, we wish to reveal how the oblateness of the primaries influence the geometry as well as shape of the Newton-Raphson basins of convergence, thus following the work of \cite{DKMP12}.

The present paper has the following structure: the most important properties of the dynamical system are presented in Section \ref{sys}. The parametric evolution of the position of the equilibrium points is investigated in Section \ref{eqpts}. The following Section contains the main numerical results, regarding the evolution of the Newton-Raphson basins of convergence. Our paper ends with Section \ref{conc}, where we emphasize the main conclusions of this work.

\section{Presentation of the dynamical system}
\label{sys}

A dimensionless, rotating, barycentric rotating system of coordinates $Oxyz$ is considered, where the two primary bodies are located on the $Ox$ axis. The primaries $P_1$ and $P_2$ have masses $m_1 = \mu$ and $m_2 = 1 - \mu$, respectively, where $\mu = m_2/(m_1 + m_2) \leq 1/2$ is the mass parameter \citep{S67}. Furthermore, the centers of both primaries are located at $(x_1, 0, 0)$ and $(x_2, 0, 0)$, where $x_1 = - \mu$ and $x_2 = 1 - \mu$. We assume that the shape of the primaries is not spherically symmetric but it resembles a spheroid. Therefore, for each primary body we introduce the corresponding oblateness coefficient $A_i$, $i = 1,2$.

According to \cite{SSR75,OV03,DM06,AS06} the time-independent effective potential function of the circular restricted-three
body problem with spheroid primaries is
\begin{equation}
\Omega(x,y,z) = \sum_{i=1}^{2} \frac{m_i}{r_i}\left(1 + \frac{A_i}{2r_i^2} - \frac{3A_i z^2}{2r_i^4}\right) + \frac{n^2}{2} \left(x^2 + y^2 \right),
\label{pot}
\end{equation}
where
\begin{align}
r_1 &= \sqrt{\left(x - x_1 \right)^2 + y^2 + z^2}, \nonumber\\
r_2 &= \sqrt{\left(x - x_2 \right)^2 + y^2 + z^2},
\label{dist}
\end{align}
are the distances of the third body from the respective primaries, while $n$ is the mean motion of the primary bodies which is defined as
\begin{equation}
n = \sqrt{1 + 3\left(A_1 + A_2 \right)/2}.
\label{mn}
\end{equation}

The equations of motion describing a test particle (third body of a negligible mass $m$, with respect to the masses of the primaries) moving under the mutual gravitational attraction of the two primaries read
\begin{equation}
\ddot{x} - 2 n \dot{y} = \frac{\partial \Omega}{\partial x}, \ \ \
\ddot{y} + 2 n \dot{x} = \frac{\partial \Omega}{\partial y}, \ \ \
\ddot{z} = \frac{\partial \Omega}{\partial z}.
\label{eqmot}
\end{equation}

The above system of differential equations admits only one integral of motion (also known as the Jacobi integral), which is
given by the following Hamiltonian
\begin{equation}
J(x,y,z,\dot{x},\dot{y},\dot{z}) = 2\Omega(x,y,z) - \left(\dot{x}^2 + \dot{y}^2 + \dot{z}^2 \right) = C,
\label{ham}
\end{equation}
where $\dot{z}$, $\dot{y}$, and $\dot{z}$ are the velocities, while $C$ is the numerical value of the Jacobi constant which is conserved.

\begin{figure}[!t]
\centering
\resizebox{0.5\hsize}{!}{\includegraphics{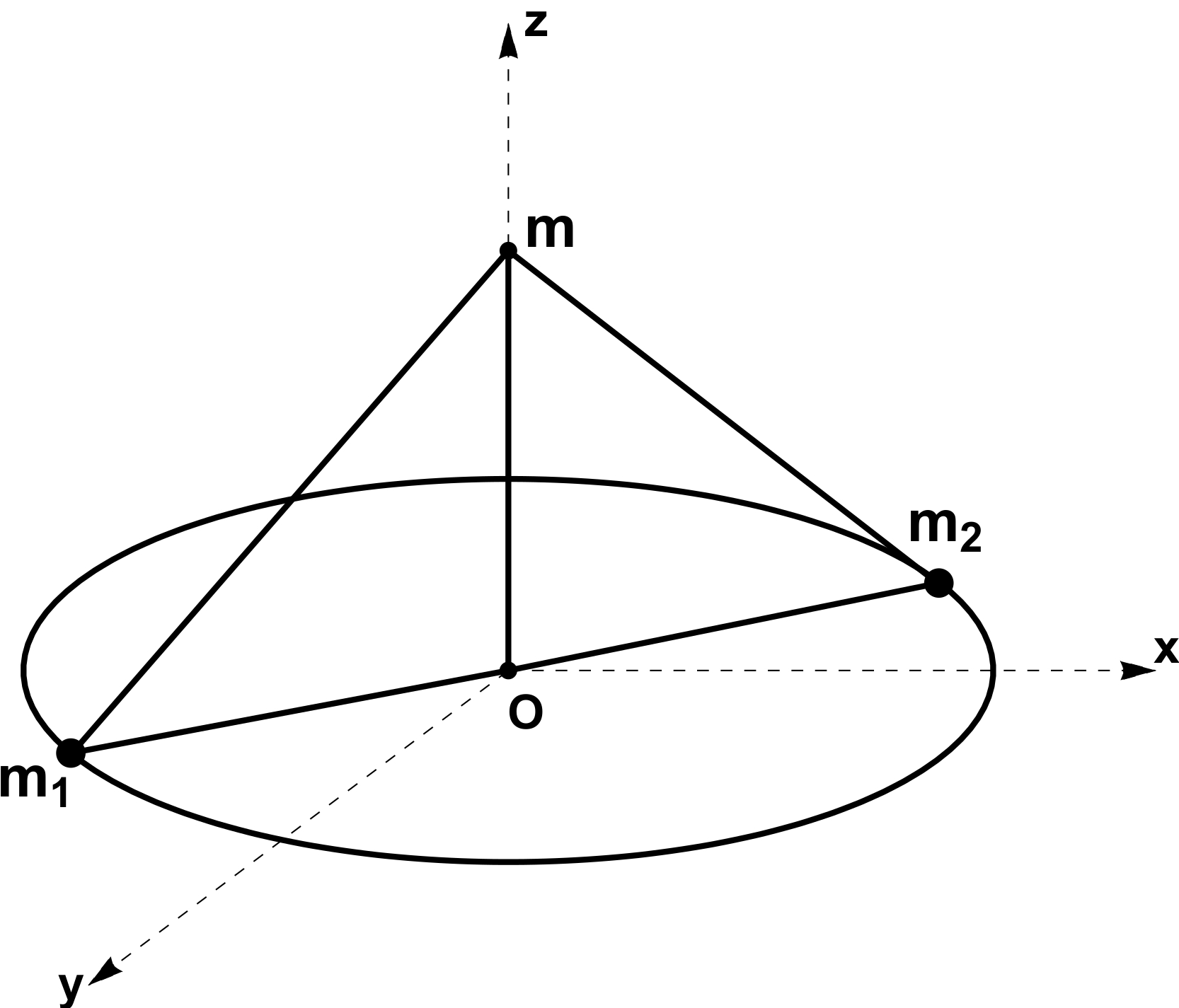}}
\caption{The configuration of the Sitnikov problem, where the two equally massed primary bodies $(m_1 = m_2 = 1/2)$ move on symmetric circular orbits.}
\label{conf}
\end{figure}

The potential function of the circular Sitnikov problem can be obtained if we set $\mu = 1/2$, $x = y = 0$, and $A_1 = A_2 = A$ in Eq. (\ref{pot}) and it equals to
\begin{equation}
\Omega(z) = \frac{1}{r} + \frac{A}{2r^3} - \frac{3Az^2}{2r^5},
\label{potz}
\end{equation}
where $r = \sqrt{z^2 + 1/4}$. It is evident that Eq. (\ref{potz}) describes the motion of a massless test particle which oscillates along a straight line which is perpendicular to the orbital $(x,y)$ plane of the two primary bodies with equal masses. In Fig. \ref{conf} we present the geometry of the Sitnikov problem.

Consequently, the equation regarding the motion of the test particle along the vertical $z$ axis has the form
\begin{equation}
\ddot{z} = - \frac{z}{r^3} - \frac{9Az}{2r^5} + \frac{15Az^3}{2r^7},
\label{eqmotz}
\end{equation}
while the corresponding Jacobi integral, for the vertical motion, becomes
\begin{equation}
J(z,\dot{z}) = 2 \Omega(z) - \dot{z}^2 = C_{z}.
\label{hamz}
\end{equation}

\section{Parametric variation of the equilibrium points}
\label{eqpts}

For locating the positions of the equilibrium points we have to set the right hand side of Eq. (\ref{eqmotz}) equal to zero as
\begin{equation}
f(\mathz;A) = - \frac{8\mathz\left(16 \mathz^4 + 8 \left(1 - 6A \right)\mathz^2 + 18 A + 1 \right)}{\left(1 + 4\mathz^2\right)^{7/2}} = 0,
\label{fza0}
\end{equation}
which is reduced to
\begin{equation}
\mathz \left(16 \mathz^4 + 8 \left(1 - 6A \right)\mathz^2 + 18 A + 1 \right) = 0.
\label{fza}
\end{equation}
Following the approach successfully used in \cite{DKMP12} (see Section 3), from now on the $z$ coordinate is considered as a complex variable and it is denoted by $\mathz$.

Looking at Eq. (\ref{fza}) we observe that the root $\mathz = 0$ is always present, regardless the value $A$ of the oblateness coefficient. This root corresponds to the inner collinear equilibrium point $L_1$ of the circular restricted three-body problem. However since the left hand side of Eq. (\ref{fza}) is a fifth order polynomial it means that there are four additional roots, given by
\begin{equation}
\mathz_i = \pm \frac{1}{2} \sqrt{6A - 1 \pm \sqrt{6A \left(6A - 5\right)}}, \ \ \ i=1,...,4.
\label{rts}
\end{equation}

The nature of these four roots strongly depends on the numerical value $A$ of the oblateness coefficient. Our analysis reveals that, along with the $\mathz = 0$ root
\begin{itemize}
  \item When $A < -1/18$ there are two real and two imaginary roots.
  \item When $A = -1/18$ there are two imaginary roots.
  \item When $A \in (-1/18,0)$ there are four imaginary roots.
  \item When $A = 0$ only the root $\mathz = 0$ exists.
  \item When $A \in (0, 5/6)$ there are four complex roots.
  \item When $A = 5/6$ there are two real roots.
  \item When $A > 5/6$ there are four real roots.
\end{itemize}
It is seen, that the values $A = \{-1/18, 0, 5/6 \}$ are in fact critical values of the oblateness coefficient, since they determine the change on the nature of the four roots.

\begin{figure}
\centering
\resizebox{0.6\hsize}{!}{\includegraphics{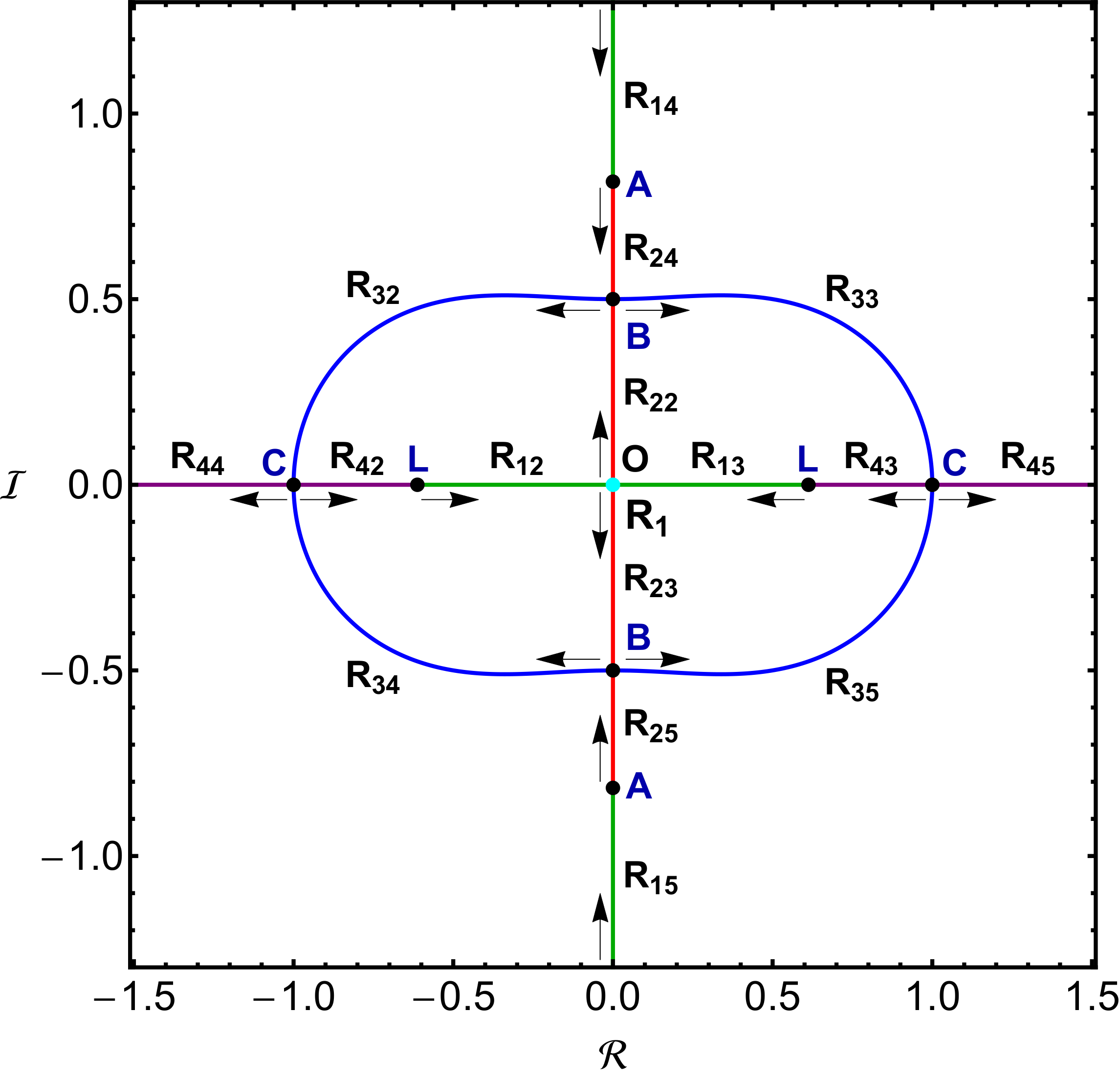}}
\caption{The space evolution of the four roots $R_{ij}$, $i,j = 1,...,4$ on the complex plane, when $A \in [-5,5]$. When $A < -1/18$ we have the roots $R_{12}$, $R_{13}$, $R_{14}$, and $R_{15}$ (green), when $A \in (-1/18, 0)$ we have the roots $R_{22}$, $R_{23}$, $R_{24}$, and $R_{25}$ (red), when $A \in (0, 5/6)$ we have the roots $R_{32}$, $R_{33}$, $R_{34}$, and $R_{35}$ (blue), while when $A > 5/6$ we have the roots $R_{42}$, $R_{43}$, $R_{44}$, and $R_{45}$ (purple). The arrows indicate the movement direction of the roots, as the value of the oblateness coefficient increases. The black dots (points A, B, and C) correspond to the three critical values of the oblateness coefficient $-1/18, 0, 5/6$, respectively, while the points L correspond to $A \to \pm \infty$. (Color figure online).}
\label{evol}
\end{figure}

It would be very interesting to determine how the positions of the four roots, on the complex plane, evolve as a function of the oblateness coefficient. Fig. \ref{evol} shows the parametric evolution of the four roots $R_{ij}$, $i,j = 1,...,4$, on the complex plane, when $A \in [-5,5]$\footnote{It should be emphasized that even for the extreme case where the equatorial radius of the non-spherical primaries takes its maximum value 1, while at the same time the polar radius takes its minimum value 0, the numerical value of the oblateness coefficient is equal to 1/5. However, in this work we shall consider much higher values of the oblateness $(A > 1/5)$ which are in fact not realistic (with no physical meaning). Nevertheless, this choice is justified if we take into account that the aim of this work is the exploration of the properties of the Newton-Raphson basins of convergence and not the actual dynamics of the circular Sitnikov three-body problem.}, with $\mathcal{R} = Re[\mathz]$ and $\mathcal{I} = Im[\mathz]$. When $A \to - \infty$ the two real roots tend to $L = \pm \sqrt{3/2}/2$, while the two imaginary roots tend to infinity. As we proceed to higher values of $A$ all four roots tend to the central region. When $A = -1/18$ the two real roots collide at the origin which increases the multiplicity of the $z = 0$ root from 1 to 3. At the same time, the two imaginary roots are located at $A = \pm \sqrt{2/3}$ on the vertical axis. As soon as $A < -1/18$ a new pair of imaginary roots emerge from the origin $(0,0)$. As the value of $A$ increases approaching 0, all four imaginary roots tend to coincide. This phenomenon occurs when $A = 0$, while the roots are exactly at $B = \pm 0.5$. For positive values of the oblateness coefficient (or in other words for oblate primaries) four complex roots emerge, one at each of the quadrants of the complex plane. As long as $A$ lies in the interval $(0, 5/6)$ the combined traces of the four complex roots create an oval shape. When $A = 5/6$ the four complex roots collide, in two pairs, on the horizontal axis, thus resulting to two real roots $C = \pm 1$ of multiplicity 2. For $A > 5/6$ two pairs of real roots emerge, while the roots of each pair move away from each other. Specifically, as $A \to \infty$ the outer roots $R_{44}$ and $R_{45}$ tend to infinity, while the roots $R_{42}$ and $R_{43}$ tend to $L = \pm \sqrt{3/2}/2$.

\section{The Newton-Raphson basins of convergence}
\label{nrb}

The easiest way of solving numerically an equation with one variable is by using the well-known Newton Raphson optimal method of second order. The corresponding iterative scheme is given by
\begin{align}
\mathz_{n+1} &= \mathz_n - \frac{f(\mathz;A)_n}{f'(\mathz;A)_n} \nonumber\\
&= \frac{12\mathz^3\left(A \left(50 - 80\mathz^2\right) + \left(1 + 4\mathz^2\right)^2 \right)}{128\mathz^6 + 48\mathz^4 - 6A \left(128\mathz^4 - 96\mathz^2 + 3 \right) - 1},
\label{nr}
\end{align}
where $\mathz_n$ is the value of the $\mathz$ at the $n$-th step of the iterative process.

The philosophy behind the Newton-Raphson method is the following: An initial complex number $\mathz = a + ib$, with $\mathcal{R} = a$ and $\mathcal{I} = b$, on the complex plane, activates the code, while the iterative procedure continues until an equilibrium point (attractor) is reached, with the desired predefined accuracy. If the particular initial condition leads to one of the roots of the system it means that the numerical method converges for that particular initial condition $(\mathcal{R}, \mathcal{I})$. At this point, it should be emphasized that in general terms the method does not converge equally well for all the available initial conditions. The sets of the initial conditions which lead to the same final state (root) compose the so-called Newton-Raphson basins of convergence or convergence domains/regions. Nevertheless, it should be clarified that the Newton-Raphson basins of convergence should not be mistaken, by no means, with the basins of attractions which are present in systems with dissipation.

\begin{figure*}[!t]
\centering
\resizebox{\hsize}{!}{\includegraphics{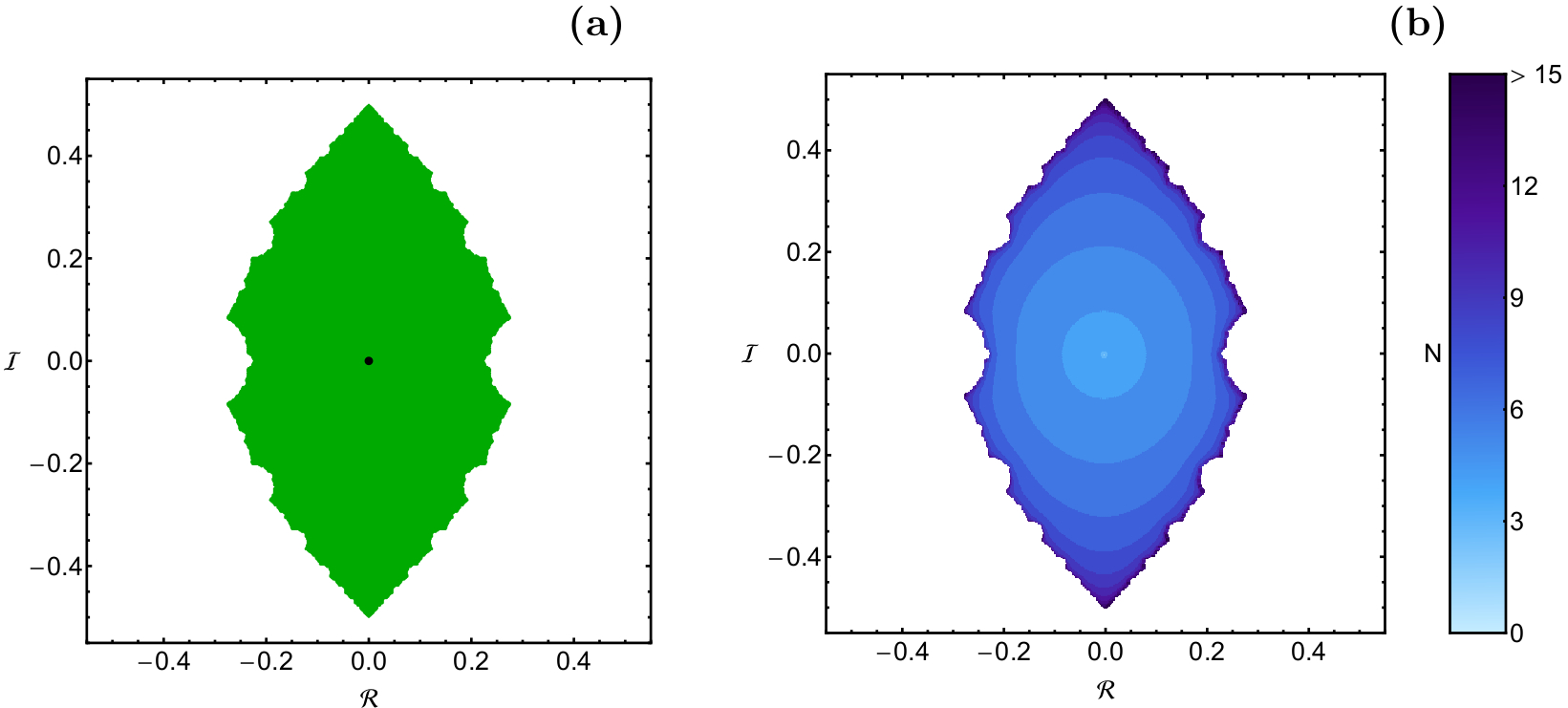}}
\caption{(a-left): The Newton-Raphson basins of convergence on the complex plane, when $A = 0$, where only one root exists. The position of the root is indicated by a black dot. The color code is as follows: $R_1$ root (green); non-converging points (white). (b-right): The distribution of the corresponding number $(N)$ of required iterations for obtaining the Newton-Raphson basins of convergence shown in panel (a). (Color figure online).}
\label{m0}
\end{figure*}

A double scan of the complex plane is performed for revealing the structures of the basins of convergence. In particular, a dense uniform grid of $1024 \times 1024$ $(\mathcal{R},\mathcal{I})$ nodes is defined which shall be used as initial conditions of the iterative scheme. The number $N$ of the iterations, required for obtaining the desired accuracy, is also monitored during the classification of the nodes. For our computations, the maximum allowed number of iterations is $N_{\rm max} = 500$. Moreover the iterations stop only when a root is reached, with accuracy of $10^{-15}$ for both real and imaginary parts.

The Newton-Raphson basins of convergence when $A = 0$ are presented in panel (a) of Fig. \ref{m0}. We see that the converging initial conditions are mainly located near the center, while they form a rhomboidal shape. In panel (b) of the same figure the distribution of the corresponding number $(N)$ of iterations required for obtaining the desired accuracy is given using tones of blue.

In the following subsections we will determine how the oblateness coefficient $A$ affects the structure of the Newton-raphson basins of convergence in the Sitnikov problem, by considering four cases regarding the nature of the roots. For the classification of the nodes on the complex plane we will use color-coded diagrams (CCDs), in which each pixel is assigned a different color, according to the final state (root) of the corresponding initial condition. Here we would like to clarify that the size of each CCD (or in other words the minimum and the maximum values of $\mathcal{R}$ and $\mathcal{I}$) is defined, in each case, in such a way so as to have a complete view of the geometry of the basins of convergence.

\begin{figure*}[!t]
\centering
\resizebox{\hsize}{!}{\includegraphics{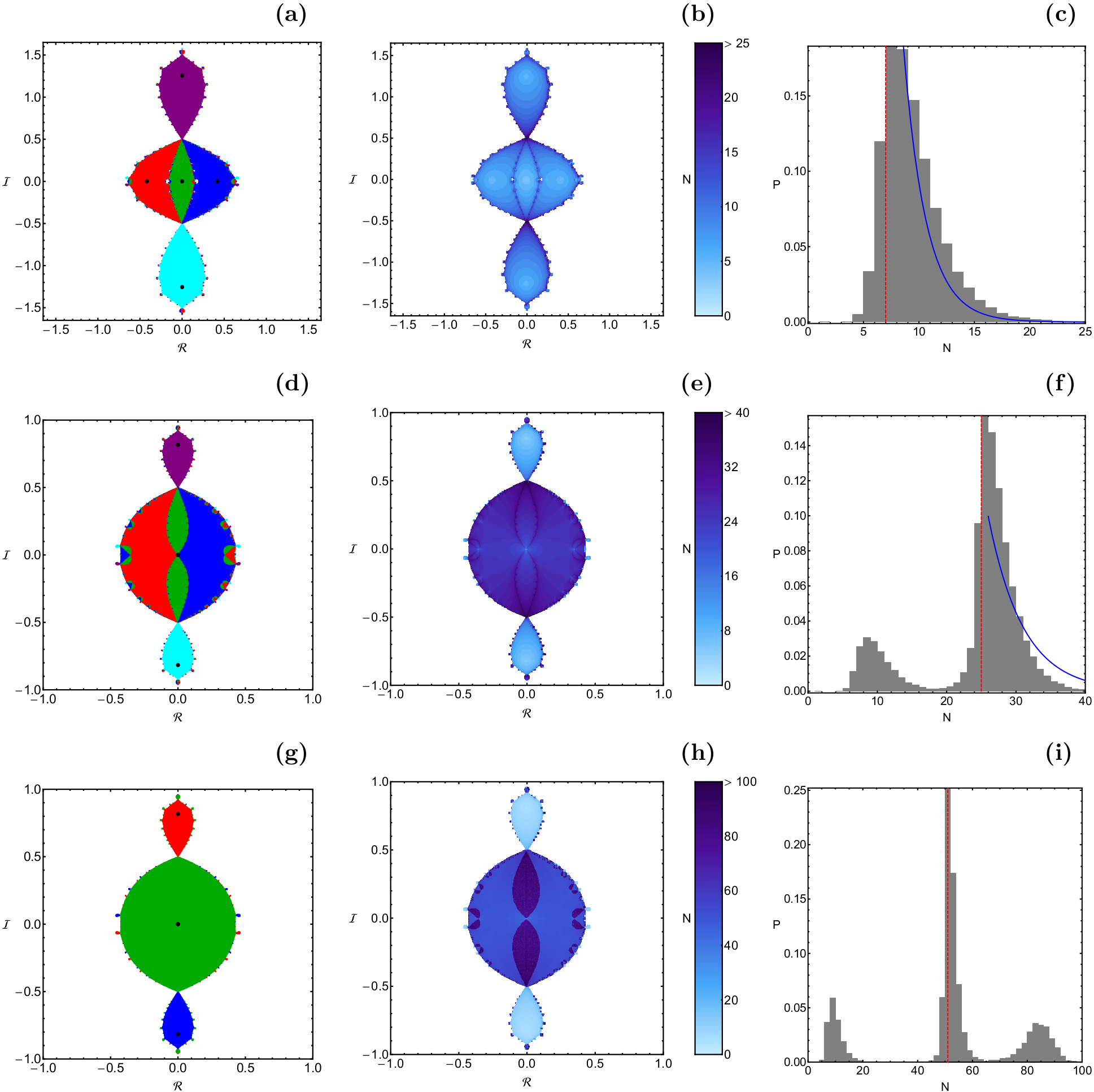}}
\caption{(First column): The Newton-Raphson basins of convergence on the complex plane for the first case, when $A \leq -1/18$. The color code, denoting the five roots, is as follows: $R_1$ (green); $R_{12}$ (red); $R_{13}$ (blue); $R_{14}$ (purple); $R_{15}$ (cyan); non-converging points (white). (Second column): The distribution of the corresponding number $N$ of required iterations for obtaining the Newton-Raphson basins of convergence. The non-converging points are shown in white. (Third column): The corresponding probability distribution of required iterations for obtaining the Newton-Raphson basins of convergence. The vertical dashed red line indicates, in each case, the most probable number $N^{*}$ of iterations. (First row): $A = -0.3$; (Second row): $A = -1/18 - 10^{-8}$; (Third row): $A = -1/18$. (Color figure online).}
\label{m1}
\end{figure*}

\begin{figure*}[!t]
\centering
\resizebox{\hsize}{!}{\includegraphics{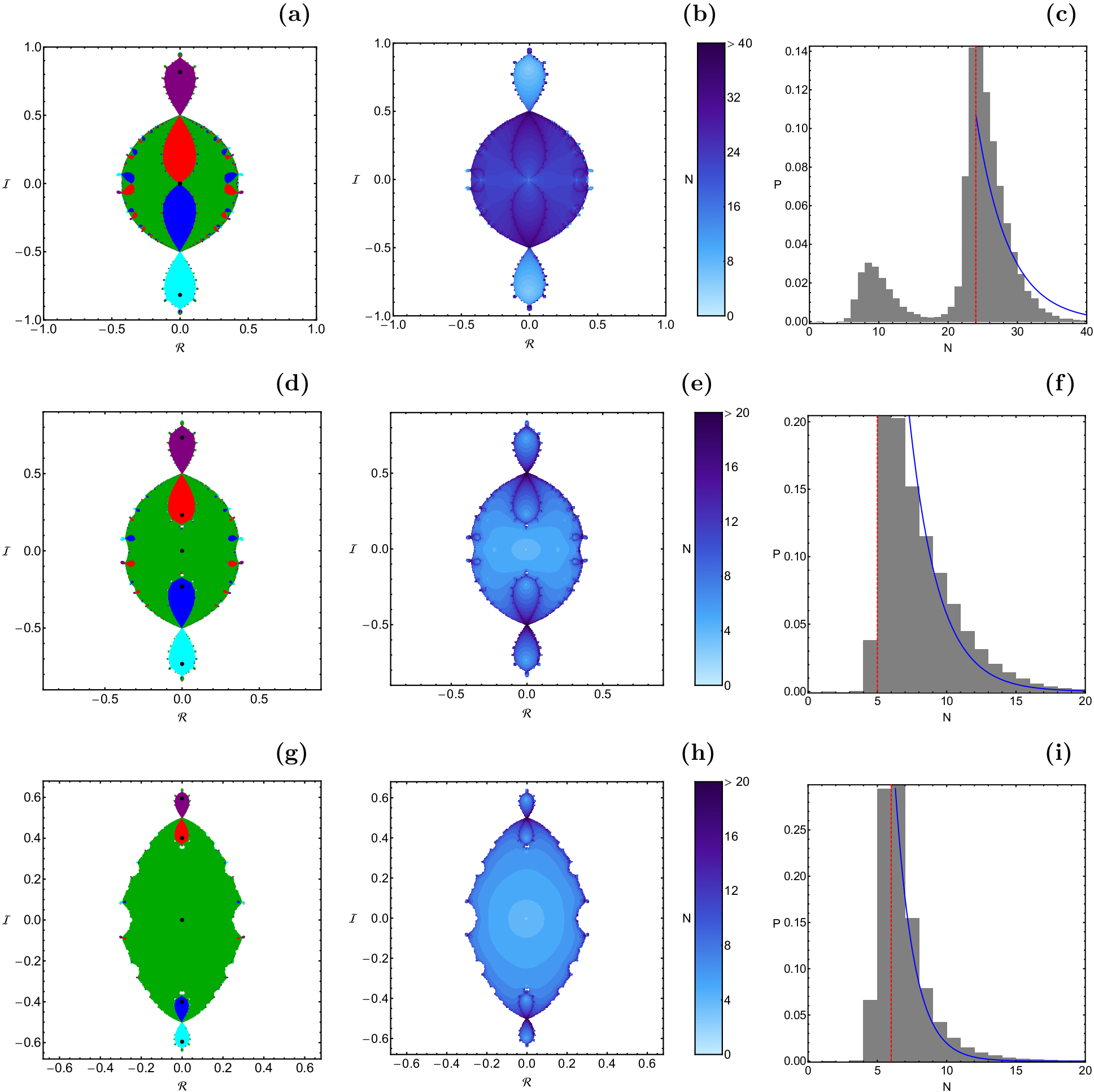}}
\caption{(First column): The Newton-Raphson basins of convergence on the complex plane for the second case, when $A \in (-1/18,0)$. The color code, denoting the five roots, is as follows: $R_1$ (green); $R_{22}$ (red); $R_{23}$ (blue); $R_{24}$ (purple); $R_{25}$ (cyan); non-converging points (white). (Second column): The distribution of the corresponding number $N$ of required iterations for obtaining the Newton-Raphson basins of convergence. The non-converging points are shown in white. (Third column): The corresponding probability distribution of required iterations for obtaining the Newton-Raphson basins of convergence. The vertical dashed red line indicates, in each case, the most probable number $N^{*}$ of iterations. (First row): $A = -1/18 + 10^{-8}$; (Second row): $A = -0.03$; (Third row): $A = -0.005$. (Color figure online).}
\label{m2}
\end{figure*}

\begin{figure*}[!t]
\centering
\resizebox{0.75\hsize}{!}{\includegraphics{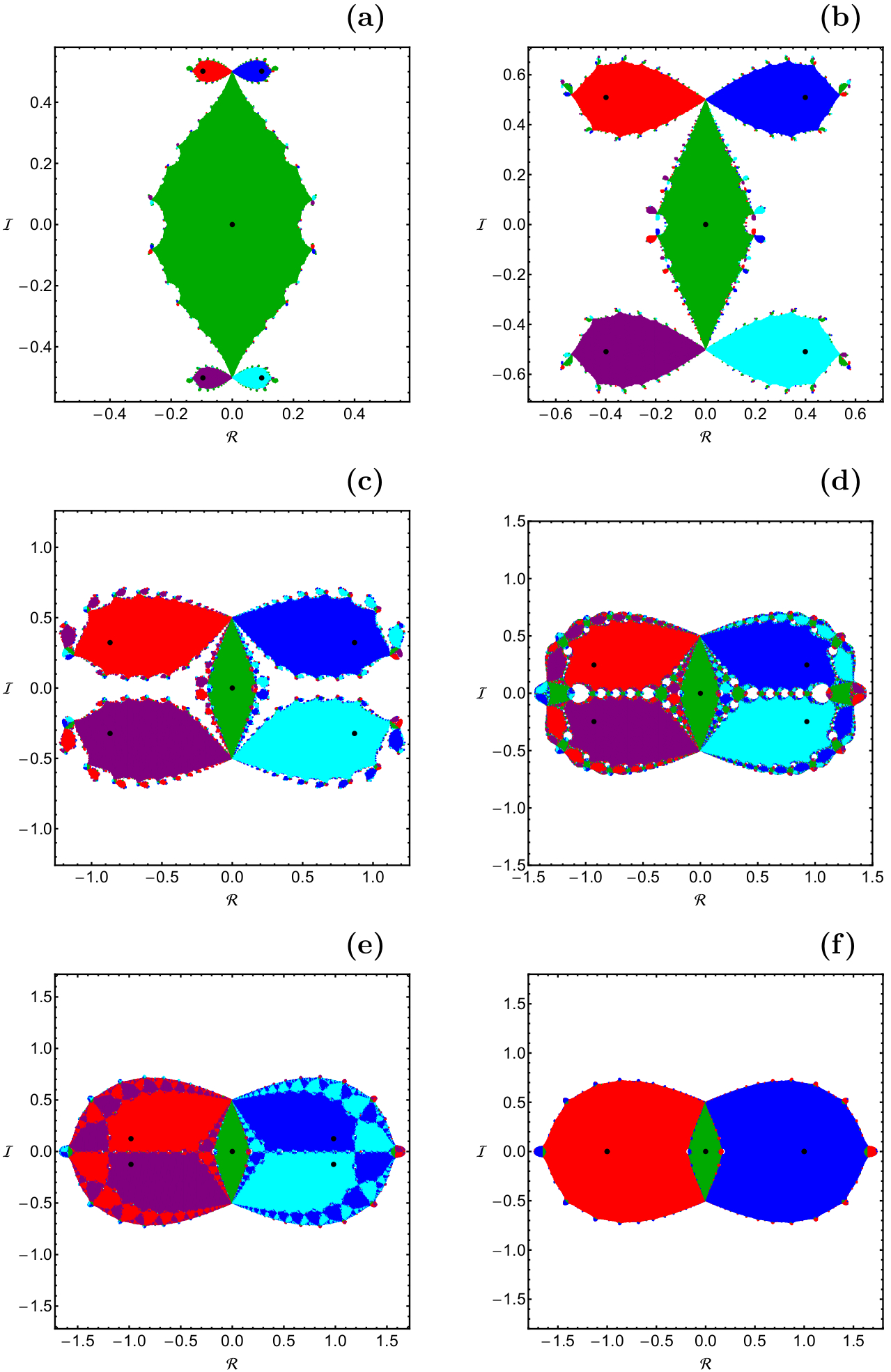}}
\caption{The Newton-Raphson basins of convergence on the complex plane for the third case, when $A \in (0,5/6]$. The color code, denoting the five roots, is as follows: $R_1$ (green); $R_{32}$ (red); $R_{33}$ (blue); $R_{34}$ (purple); $R_{35}$ (cyan); non-converging points (white). (a): $A = 0.005$; (b): $A = 0.1$; (c): $A = 0.6$; (d): $A = 0.7$; (e): $A = 0.8$; (f): $A = 5/6$. (Color figure online).}
\label{c3}
\end{figure*}

\begin{figure*}[!t]
\centering
\resizebox{0.85\hsize}{!}{\includegraphics{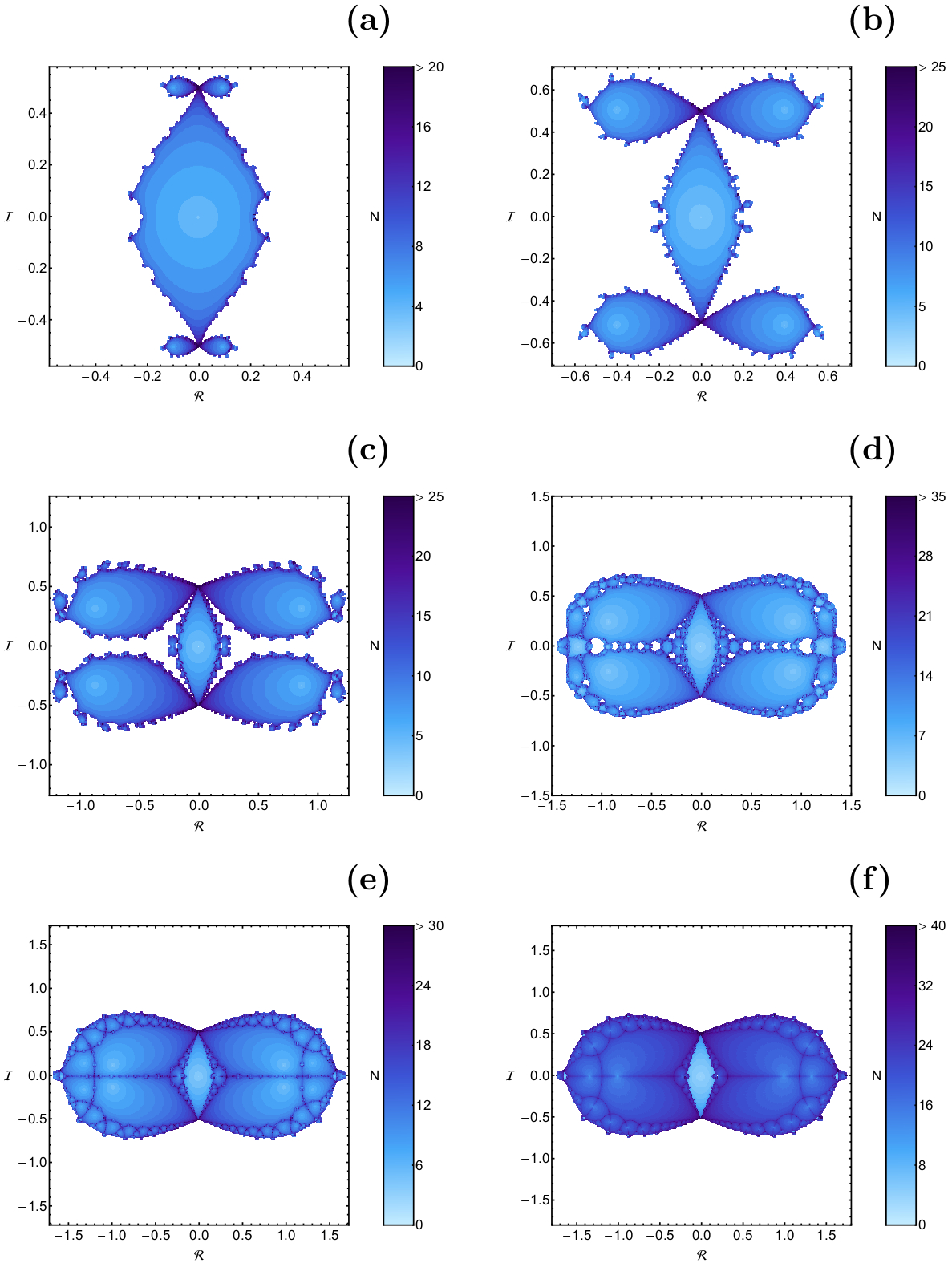}}
\caption{The distribution of the corresponding number $N$ of required iterations for obtaining the Newton-Raphson basins of convergence, shown in Fig. \ref{c3}. The non-converging points are shown in white. (Color figure online).}
\label{n3}
\end{figure*}

\begin{figure*}[!t]
\centering
\resizebox{0.75\hsize}{!}{\includegraphics{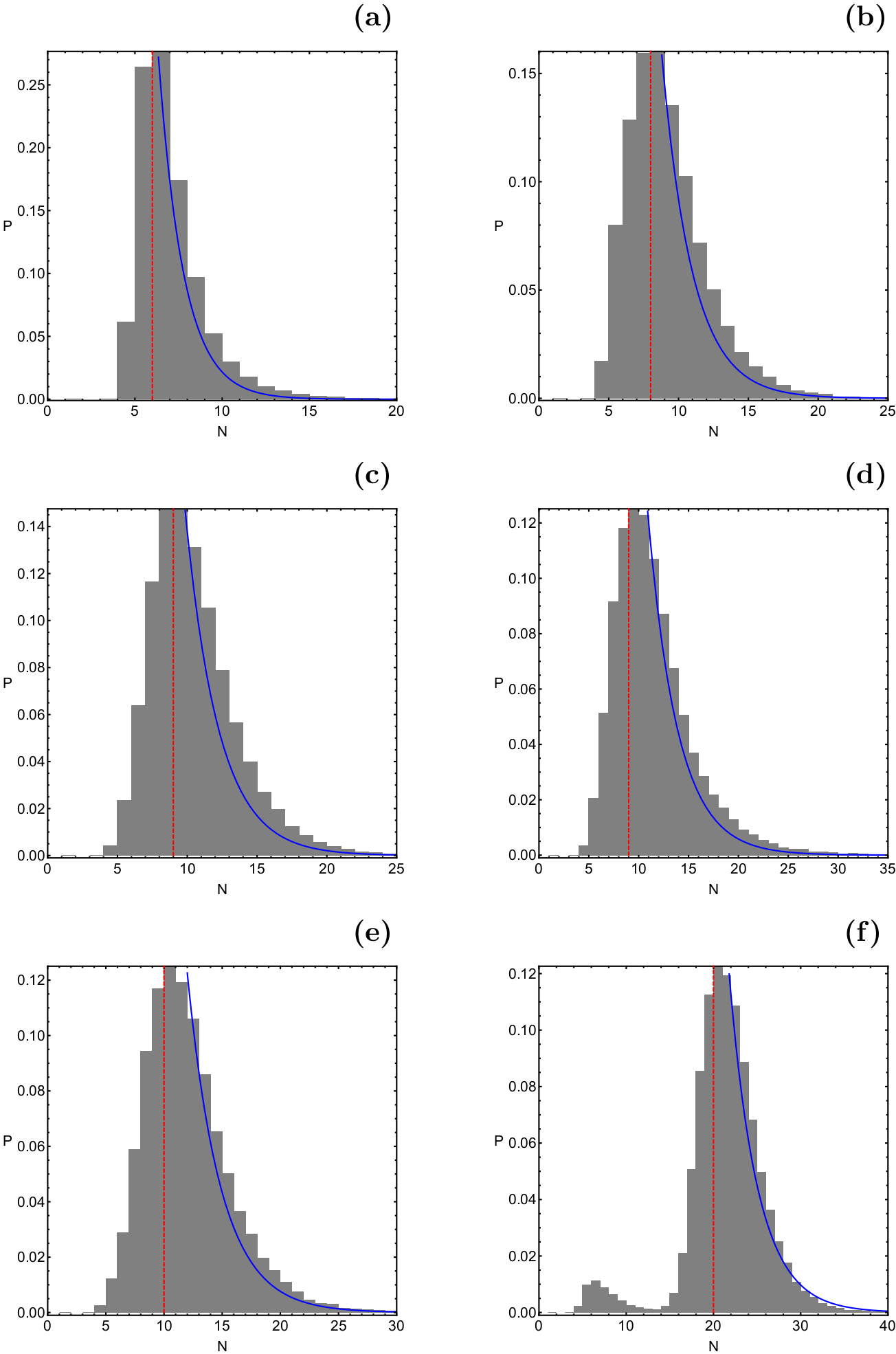}}
\caption{The corresponding probability distribution of required iterations for obtaining the Newton-Raphson basins of convergence, shown in Fig. \ref{c3}. The vertical dashed red line indicates, in each case, the most probable number $N^{*}$ of iterations. (Color figure online).}
\label{p3}
\end{figure*}

\begin{figure*}[!t]
\centering
\resizebox{\hsize}{!}{\includegraphics{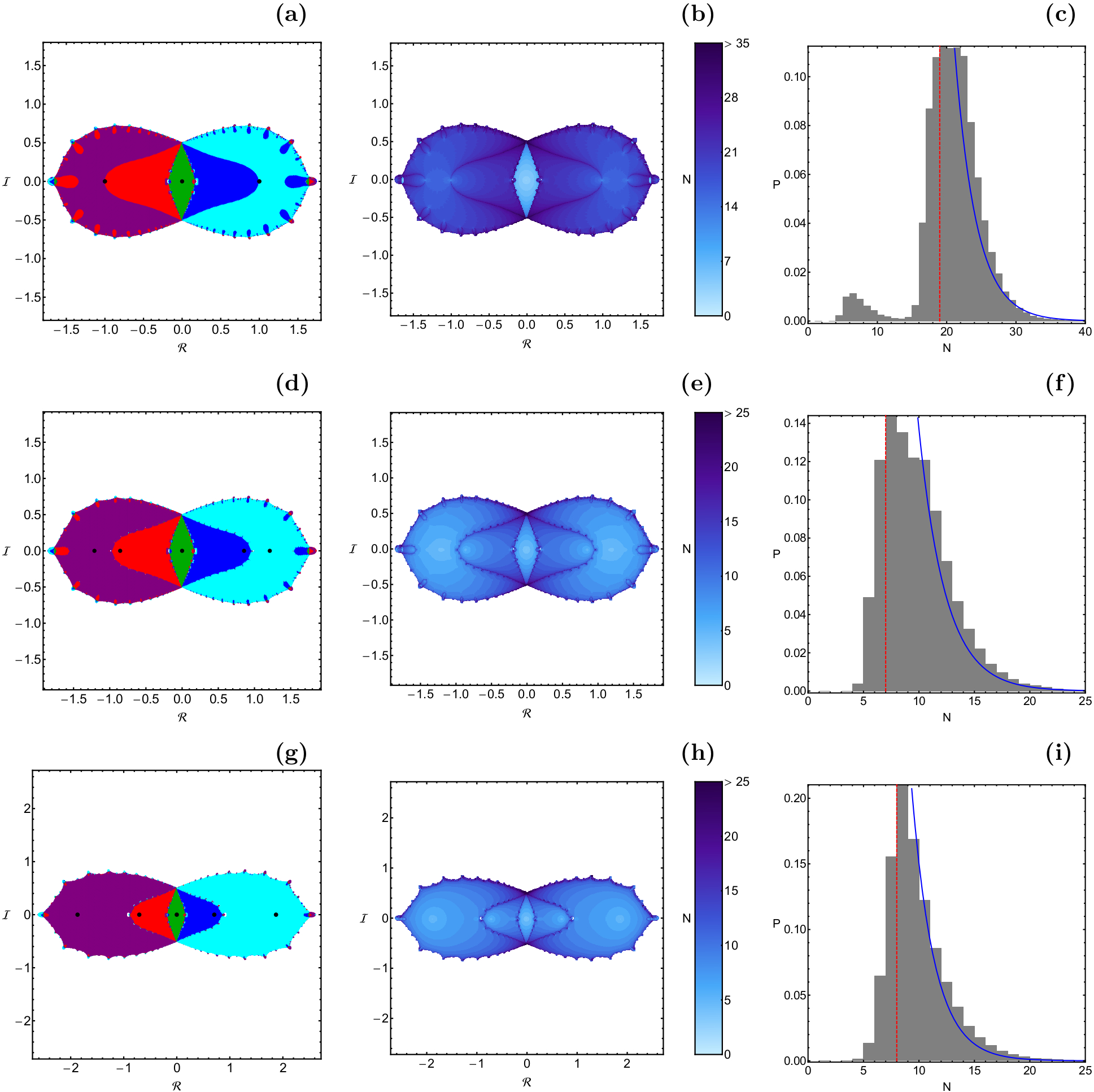}}
\caption{(First column): The Newton-Raphson basins of convergence on the complex plane for the fourth case, when $A > 5/6$. The color code, denoting the five roots, is as follows: $R_1$ (green); $R_{42}$ (red); $R_{43}$ (blue); $R_{44}$ (purple); $R_{45}$ (cyan); non-converging points (white). (Second column): The distribution of the corresponding number $N$ of required iterations for obtaining the Newton-Raphson basins of convergence. The non-converging points are shown in white. (Third column): The corresponding probability distribution of required iterations for obtaining the Newton-Raphson basins of convergence. The vertical dashed red line indicates, in each case, the most probable number $N^{*}$ of iterations. (First row): $A = 5/6 + 10^{-8}$; (Second row): $A = 0.9$; (Third row): $A = 1.5$. (Color figure online).}
\label{m4}
\end{figure*}

\begin{figure*}[!t]
\centering
\resizebox{0.70\hsize}{!}{\includegraphics{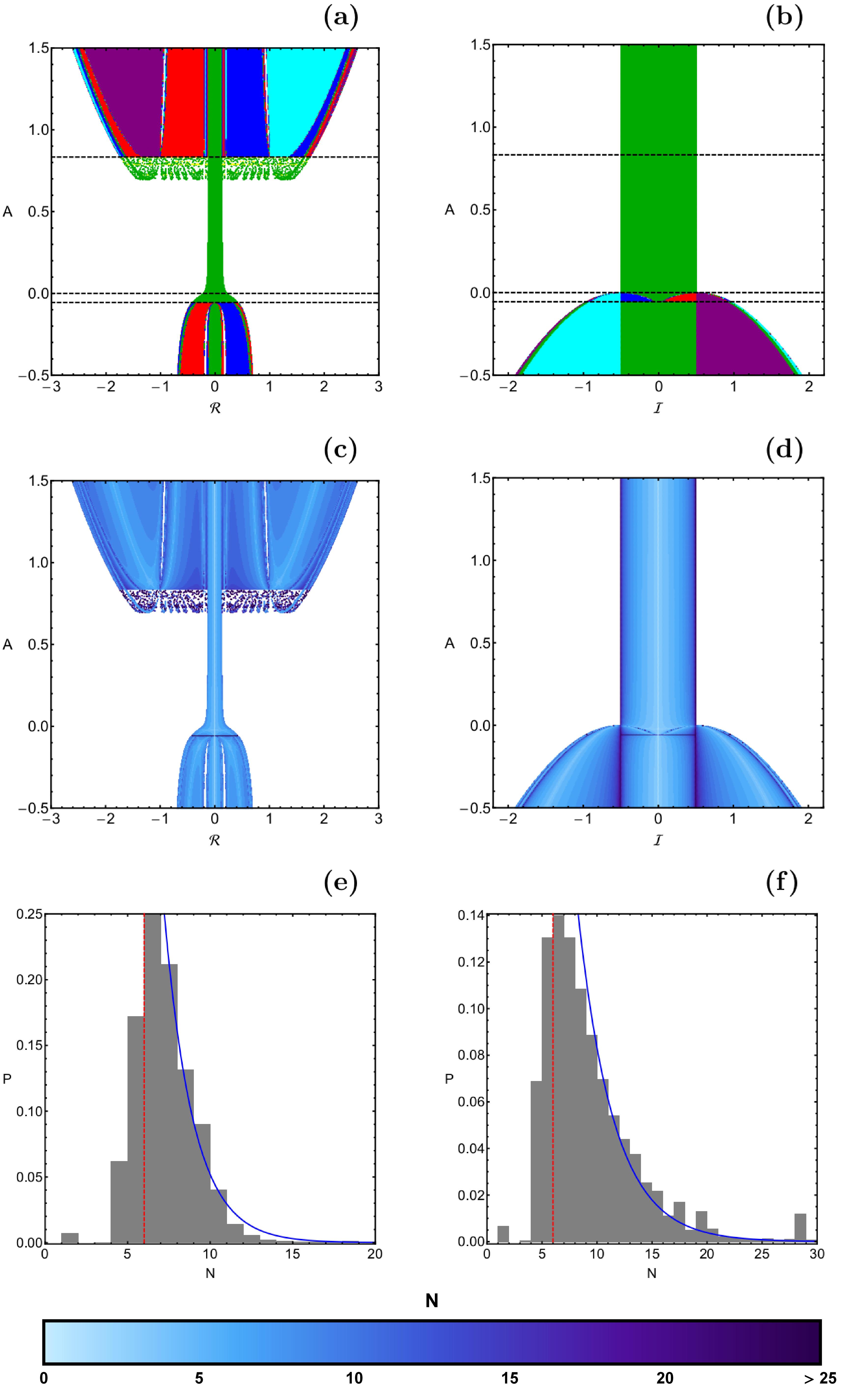}}
\caption{(First row): The Newton-Raphson basins of convergence on the (a-left): $(\mathcal{R},A)$ and (b-right): $(\mathcal{I},A)$ plane, where $A \in [-0.5,1.5]$. The color code, denoting the five roots, is as follows: $R_1$ (green); $R_{i2}$ (red); $R_{i3}$ (blue); $R_{i4}$ (purple); $R_{i5}$ (cyan); non-converging points (white), with $i = 1,...,4$. (Second row): The distribution of the corresponding number $N$ of required iterations for obtaining the Newton-Raphson basins of convergence. The non-converging points are shown in white. (Third row): The corresponding probability distribution of required iterations for obtaining the Newton-Raphson basins of convergence. The vertical dashed red line indicates, in each case, the most probable number $N^{*}$ of iterations. (Color figure online).}
\label{ria}
\end{figure*}

\subsection{Case I: $A \leq -1/18$}
\label{ss1}

We begin with the first case, where the equation $f(\mathz;A) = 0$ has, apart from the $\mathz = 0$ root, two real and two imaginary roots. The Newton-Raphson basins of convergence on the complex plane, for three values of the oblateness coefficient, are illustrated in the first column of Fig. \ref{m1}. It is seen that in all cases the area of all the types of the basins of convergence is finite. On the contrary, outside the convergence regions the vast majority of the complex plane is covered by initial conditions which do not converge to any of the five roots (white regions). Additional numerical calculations indicate that for all these non-converging initial conditions the Newton-Raphson iterative procedure lead to infinity (which numerically equals to extremely large numbers).

In the second column of Fig. \ref{m1} we present the corresponding number $N$ of iterations, using tones of blue, while the corresponding probability distribution of the required iterations is given in the third column of the same figure. The definition of the probability $P$ is the following: if $N_0$ complex initial conditions $(\mathcal{R},\mathcal{I})$ converge, after $N$ iterations, to one of the roots then $P = N_0/N_t$, where $N_t$ is the total number of nodes in every CCD. Moreover, in all plots the tails of the histograms extend so as to cover 97\% of the corresponding distributions of iterations. The vertical, red, dashed line in the probability histograms denote the most probable number $N^{*}$ of iterations. The blue lines in the histograms of Fig. \ref{m1} indicate the best fit to the right-hand side $N > N^{*}$ of them (more details are given in subsection \ref{geno}).

The diagrams shown in the second and third column of Fig. \ref{m1} allow us to extract additional information regarding the basins of convergence. Indeed, in panel (b), where $A = -0.3$, we observe that the initial conditions inside the several basins converge relatively fast (within the first 10 to 12 iterations) to one of the roots. On the other hand, all the initial conditions in the vicinity of the basin boundaries require more than 20 iterations in order to converge to one of the roots. In reality the regions in the vicinity of the basin boundaries are highly fractal\footnote{By the term fractal we simply mean that the particular area has a fractal-like geometry, without conducting any additional calculations for computing the fractal dimension as in \cite{AVS01,AVS09}.}, which implies that the final state (root) of an initial condition inside this area is highly sensitive. More precisely, even the slightest change of the initial conditions automatically leads to a completely different root, which is a classical indication of chaos. Therefore for the initial conditions in the basin boundaries it is almost impossible to predict their final states (roots).

In panel (f) of Fig. \ref{m1}, where $A = -1/18 - 10^{-8}$, one can identify two peaks on the histogram. This is because the corresponding distribution diagram, shown in panel (e) of the same figure, suggests that the two lobes, corresponding to roots $R_{14}$ and $R_{15}$, are composed of fast converging initial conditions, while the central basins, corresponding to roots $R_1$, $R{12}$, and $R_{13}$ are composed of slow converging initial conditions. When $A = -1/18$ we see in panel (g) of Fig. \ref{m1} that the central oval region is entirely populated by initial conditions which lead to root $\mathz = 0$ $(R_1)$. However, the corresponding distribution diagram, shown in panel (h) of the same figure, reveals some interesting hidden patterns inside this central region. In particular, there are two additional lobes as well as some minor structures at the borders of the oval region. In fact we may say that the distribution diagram suggests that the converging initial conditions in this case can be divided into three categories: (i) fast converging, (ii) slow converging, and (iii) very slow converging points. The corresponding probability distribution diagram, shown in panel (i), is in complete agreement, showing the three anticipated peaks.

\subsection{Case II: $A \in (-1/18,0)$}
\label{ss2}

The next case under consideration involves the scenario where there are four pure imaginary roots, along with the $\mathz = 0$ root. In the first column of Fig. \ref{m2} we present the Newton-Raphson basins of convergence for three values of the oblateness coefficient. As we proceed to higher values of $A$ the main changes, regarding the geometry of the convergence areas, are the following:
\begin{itemize}
  \item The extent of the four lobes, corresponding to roots $R_{22}$, $R_{23}$, $R_{24}$, and $R_{25}$, constantly decreases.
  \item The geometry of the central basin, corresponding to root $R_1$ changes from oval to rhomboidal.
  \item The basin boundaries of the central basin become more smooth, while at the same time all the fractal areas in the vicinity of the basin boundaries are heavily confined.
\end{itemize}

The second and third column of Fig. \ref{m2} contain the corresponding number $N$ of iterations, and the probability distribution of the required iterations, respectively. In panel (b), where $A = -1/18 + 10^{-8}$, we observe that the required number of iterations for the initial conditions which converge to roots $R_{1}$, $R_{22}$, and $R_{23}$, is more than twice the required number, regarding the initial conditions which lead to roots $R_{24}$, and $R_{25}$. Indeed, in the corresponding probability distribution of panel (c) we can distinguish the two peaks, which indicate that the converging initial conditions are divided into fast and slow converging.

\subsection{Case III: $A \in (0,5/6]$}
\label{ss3}

We continue with the case where the equation $f(\mathz;A) = 0$ has four complex roots, along with the $R_1$ root $\mathz = 0$. The evolution of the geometry of the Newton-Raphson basins of convergence is depicted in Fig. \ref{c3}, where we present six CCDs for six values of the oblateness coefficient. As the value of $A$ increases the following phenomena take place:
\begin{itemize}
  \item The area of the central region, corresponding to root $R_1$ decreases, while the area of the four lobes constantly increases.
  \item The orientation of the four lobes is now parallel to the horizontal axis, while in the two previous cases (which correspond to prolate $A < 0$ primaries) the lobes where in vertical orientation with respect to the horizontal axis.
  \item As we tend to the third critical value of the oblateness coefficient $(A = 5/6)$, the four lobes merge together, on the horizontal axis, thus creating unified basins of convergence.
\end{itemize}

The distribution of the required iterations $N$ is illustrated in Fig. \ref{n3}, while the probability distribution of the required iterations is given in Fig. \ref{p3}. In panel (f) of Fig. \ref{n3} it is seen that the required iterations for reaching the two real roots $\pm 1$ are relatively high, with respect to the required iteration for reaching the central root $\mathz = 0$. In particular, according to panel (f) of Fig. \ref{p3} the initial conditions which converge to root $R_1$ need about 7 iteration, while the initial conditions which converge to the two real roots need, approximately, more than 20 iterations, for obtaining the desired accuracy.

\subsection{Case IV: $A > 5/6$}
\label{ss4}

Our exploration ends with the case where there are four real roots, along with the $\mathz = 0$ root. The Newton-Raphson basins of convergence, for three values of the oblateness coefficient $A$, are depicted in the first column of Fig. \ref{m4}. The corresponding number $N$ of iterations, and the probability distribution of the required iterations are give in the second and third column of Fig. \ref{m4}, respectively. In panel (b), where $A = 5/6 + 10^{-8}$, it is seen that the vast majority of the initial conditions converge to one of the roots $R_{42}$, $R_{43}$, $R_{44}$, and $R_{45}$, only after about 18 iterations. On the other hand, all the initial conditions, which form the central basin, need no more 10 iterations to converge to the root $R_1$. With increasing value of the oblateness coefficient $A$ the geometry of the complex plane changes as follows:
\begin{itemize}
  \item The extent of the basins of convergence, corresponding to roots $R_1$, $R_{42}$, and $R_{43}$, decreases.
  \item The area of the convergence regions, corresponding to roots $R_{44}$ and $R_{45}$, increases.
  \item The boundaries of the basins, corresponding to roots $R_{44}$ and $R_{45}$, become more smooth and all the small fractal regions are significantly been reduced.
\end{itemize}

\subsection{An overview analysis}
\label{geno}

The color-coded convergence diagrams on the complex plane, presented earlier in subsections \ref{ss1}, \ref{ss2}, \ref{ss3}, and \ref{ss4} provide sufficient information regarding the attracting domains, however for only a fixed value of the oblateness coefficient $A$. In order to overcome this handicap we can define a new type of distribution of initial conditions which will allow us to scan a continuous spectrum of $A$ values, rather than few discrete levels. The most interesting configuration is to set either the real part or the imaginary part equal to zero, while the value of the oblateness coefficient will vary in the interval $[-0.5,1.5]$. This technique allows us to construct, once more, a two-dimensional plane in which the $\mathcal{R}$ or the $\mathcal{I}$ is the abscissa, while the value of $A$ is always the ordinate. The first row of Fig. \ref{ria} shows the basins of convergence on the $(\mathcal{R},A)$ and $(\mathcal{I},A)$ planes, while the distribution of the corresponding number $N$ of required iterations and the probability distributions are given in the second and third column of Fig. \ref{ria}, respectively. In panels (a) and (b) of Fig. \ref{ria} it can be seen very clearly how the convergence properties of the system change, as a function of the oblateness coefficient.

Additional interesting information could be extracted from the probability distributions of iterations presented in the second row of Fig. \ref{ria}. In particular, it would be very interesting to try to obtain the best fit of the tails\footnote{By the term ``tails" of the distributions we refer to the right-hand side of the histograms, that is, for $N > N^{*}$.} of the distributions. For fitting the tails of the histograms, we used the Laplace distribution, which is the most natural choice, since this type of distribution is very common in systems displaying transient chaos (e.g., \cite{ML01,SASL06,SS08}). Our calculations strongly indicate that in the vast majority of the cases the Laplace distribution is the best fit to our data. The only case where the Laplace distribution fails to properly fit the corresponding numerical data is the cases corresponding to $A = -1/18$, where there is an additional peak after the major peak (see panel (i) of Fig. \ref{m1}).

The probability density function (PDF) of the Laplace distribution is given by
\begin{equation}
P(N | a,b) = \frac{1}{2b}
 \begin{cases}
      \exp\left(- \frac{a - N}{b} \right), & \text{if } N < a \\
      \exp\left(- \frac{N - a}{b} \right), & \text{if } N \geq a
 \end{cases},
\label{pdf}
\end{equation}
where $a$ is the location parameter, while $b > 0$, is the diversity. In our case we are interested only for the $x \geq a$ part of the distribution function.

In Table \ref{t1} we present the values of the location parameter $a$ and the diversity $b$, as they have been obtained through the best fit, for all cases discussed in the previous subsections. One may observe that for most of the cases the location parameter $a$ is very close to the most probable number $N^{*}$ of iterations, while in some cases these two quantities coincide.

\begin{table}[!ht]
\begin{center}
   \caption{The values of the location parameter $a$ and the diversity $b$, related to the most probable number $N^{*}$ of iterations, for all the studied cases shown earlier in the CCDs.}
   \label{t1}
   \setlength{\tabcolsep}{10pt}
   \begin{tabular}{@{}lrrrr}
      \hline
      Figure & $A$ & $N^{*}$ & $a$ & $b$ \\
      \hline
      \ref{m0}a &             0 &  6 & $N^{*}$     & 1.15 \\
      \hline
      \ref{m1}c &           -0.30 &  7 & $N^{*} + 1$ & 2.03 \\
      \ref{m1}f & $-1/18-10^{-8}$ & 25 & $N^{*} + 1$ & 5.01 \\
      \ref{m1}i &           -1/18 & 51             - &    - \\
      \hline
      \ref{m2}c & $-1/18+10^{-8}$ & 24 & $N^{*}$     & 4.66 \\
      \ref{m2}f &           -0.03 &  5 & $N^{*} + 2$ & 2.12 \\
      \ref{m2}i &          -0.005 &  6 & $N^{*}$     & 1.35 \\
      \hline
      \ref{p3}a &           0.005 &  6 & $N^{*}$     & 1.44 \\
      \ref{p3}b &             0.1 &  8 & $N^{*}$     & 2.19 \\
      \ref{p3}c &             0.6 &  9 & $N^{*}$     & 2.38 \\
      \ref{p3}d &             0.7 &  9 & $N^{*} + 1$ & 2.97 \\
      \ref{p3}e &             0.8 & 10 & $N^{*} + 1$ & 2.88 \\
      \ref{p3}f &             5/6 & 20 & $N^{*} + 1$ & 3.26 \\
      \hline
      \ref{m4}c &   $5/6+10^{-8}$ & 19 & $N^{*} + 1$ & 3.09 \\
      \ref{m4}f &             0.9 &  7 & $N^{*} + 2$ & 2.41 \\
      \ref{m4}i &             1.5 &  8 & $N^{*} + 1$ & 2.03 \\
      \hline
      \ref{ria}c &              - &  6 & $N^{*} + 1$ & 1.76 \\
      \ref{ria}d &              - &  6 & $N^{*} + 2$ & 3.25 \\
      \hline
   \end{tabular}
\end{center}
\end{table}

\section{Concluding remarks}
\label{conc}

We numerically explored the basins of convergence in the Sitnikov three-body problem with non-spherical primaries. More precisely, we demonstrated how the oblateness coefficient $A$ influences the position of the roots on the complex plane. The Newton-raphson iterative scheme was used for revealing the corresponding basins of convergence on the complex plane. These convergence domains play a significant role, since they explain how each point of the complex plane is attracted by the equilibrium points of the system, which act, in a way, as attractors. We managed to monitor how the Newton-Raphson basins of convergence evolve as a function of the oblateness coefficient. Another important aspect of this work was the relation between the basins of convergence and the corresponding number of required iterations and the respective probability distributions.

To our knowledge this is the first time that the Newton-Raphson basins of convergence in the Sitnikov problem are numerically investigated in such a systematic and thorough manner. On this basis, the presented results are novel and this is exactly the contribution of our work.

The following list contains the most important conclusions of our numerical analysis.
\begin{enumerate}
  \item Real roots are possible for both prolate $(A <0)$ and oblate $(A > 0)$ configurations of the primaries, while on the other hand, imaginary roots are possible only for prolate primaries. Complex roots exist only when the oblateness coefficient lies in the interval $(0, 5/6)$.
  \item It was found that all the basins of convergence, corresponding to all five roots, have finite area, regardless the value of the oblateness coefficient.
  \item Our numerical analysis indicates that the vast majority of the complex plane is covered by initial conditions which do not converge to any of the five roots. Furthermore, additional computations revealed that for all these initial conditions the Newton-Raphson iterator lead to extremely large complex numbers (either real or imaginary), which implies that these initial conditions tend to infinity.
  \item Near the critical values of the oblateness coefficient we identified several types of converging areas for which the corresponding number of required iterations is relatively high, with respect to near by basins of other roots. We suspect that this phenomenon is inextricably linked with the fact that near these critical points the dynamics of the system, such as the total number of the equilibrium points (roots), changes.
  \item The Newton-Raphson method was found to converge very fast $(0 \leq N < 10)$ for initial conditions close to the roots, fast $(10 \leq N < 15)$ and slow $(15 \leq N < 30)$ for initial conditions that complement the central regions of the very fast convergence, and very slow $(N \geq 30)$ for initial conditions of dispersed points lying either in the vicinity of the basin boundaries, or between the dense regions of the roots.
\end{enumerate}

A double precision numerical code, written in standard \verb!FORTRAN 77! \cite{PTVF92}, was used for the classification of the initial conditions into the different types of basins. In addition, for all the graphical illustration of the paper we used the latest version 11.2 of Mathematica$^{\circledR}$ \cite{W03}. Using an Intel$^{\circledR}$ Quad-Core\textsuperscript{TM} i7 2.4 GHz PC the required CPU time, for the classification of each set of initial conditions, was about 5 minutes.

In the future, it would be very interesting to use other types of iterative schemes and compare the similarities as well as the differences on the corresponding basins of convergence. In particular, using iterative methods of higher order, with respect to the classical Newton-Raphson method, would be an ideal starting point, for demystifying the secrets of this active field of research.

\section*{Acknowledgments}

The authors would like to express their warmest thanks to the anonymous referee for the careful reading of the manuscript and for all the apt suggestions and comments which allowed us to improve both the quality as well as the clarity of the paper.

\footnotesize
\section*{Compliance with Ethical Standards}

\begin{itemize}
  \item Funding: The authors state that they have not received any research grants.
  \item Conflict of interest: The authors declare that they have no conflict of interest.
\end{itemize}

\end{document}